# Enhancement of Superconducting Properties of Polycrystalline CaKFe$_4$As$_4$ by High-Pressure Growth

Manasa Manasa[1], Mohammad Azam[1], Tatiana Zajarniuk[2], Ryszard Diduszko[3], Tomasz Cetner[1], Andrzej Morawski[1], Andrzej Wiśniewski[2], Shiv J. Singh[1*]

[1]Institute of High Pressure Physics (IHPP), Polish Academy of Sciences, Sokołowska 29/37, 01-142 Warsaw, Poland.

[2]Institute of Physics, Polish Academy of Sciences, aleja Lotników 32/46, 02-668 Warsaw, Poland.

[3]Łukasiewicz Research Network Institute of Microelectronics and Photonics, Aleja Lotników 32/46, 02-668 Warsaw, Poland

*Corresponding Author:  sjs@unipress.waw.pl

*Abstract*

High-pressure growth is a unique method to improve the sample quality and size. Here, we have used the high gas pressure and high-temperature synthesis (HP-HTS) method to grow CaKFe$_4$As$_4$ (1144) bulks and investigated their superconducting properties using structural, microstructural, transport, and magnetic studies. The microstructural analysis demonstrates that 1144 samples prepared by HP-HTS have improved the sample density and grain connectivity. The transition temperature ($T_c^{onset}$) of 1144 bulks prepared by HP-HTS is increased up to 35.2 K with a transition width ($\Delta T$) of 1 K, which is remarkably comparable to the reported 1144 single crystal. Additionally, the critical current density ($J_c$) is enhanced by almost one order of magnitude compared with the parent compound prepared by the conventional synthesis process at ambient pressure (CSP), which could be attributed to the improved sample density and effective pinning centers. Our study demonstrates that the sample quality and superconducting properties of various iron-based superconductors can be enhanced by applying the HP-HTS approach, and further research is demanded in this direction.

## I. INTRODUCTION

Iron-based superconductors (FBS) are a high $T_c$ material [1],[2] with high transition temperature ($T_c$) up to 58 K [3] and a high upper critical field ($H_{c2}$) up to 100 T [4-5]. More than 100 compounds have been discovered that can be categorized in six families based on the crystal structure of the parent compound [1][2][6]: *RE*FeAsO (*RE*: Rare earth) (1111), *AE*Fe$_2$As$_2$ (*AE* = Ba, Ca) (122), FeSe (11), *A*FeAs (*A* = Li, Na) (111), *AeA*1144 (*Ae* = Ca, *A* = K) and thick perovskite oxide blocking layers such as Sr$_4$Sc$_2$O$_6$Fe$_2$P$_2$ (42622). Most of the compounds depict the superconductivity by chemical doping, but there also exist some stoichiometry compounds such as LiFeAs (111), FeSe (11), and CaKFe$_4$As$_4$ (1144). CaKFe$_4$As$_4$ [7] is one of the most promising stoichiometric compounds belonging to the 1144 family that shows the highest transition temperature up to 35-36 K without any chemical doping effect [8],[9]. It is well known that the control of chemical doping during the growth process is a difficult job, which generally produces inhomogeneity inside the samples and many impurity phases [2]. These impurity phases generally reduce the superconducting properties. Therefore, it may be more beneficial to use the stoichiometric superconducting compound, meaning that the control of chemical doping contents during sample growth need not be a concern [7],[9],[10],[11].

1144 family was discovered in 2016 and has a crystal structure almost similar to the 122 family. In the last six years, studies have proved that these materials have the highest $T_c$ of 35 K and a very high critical current density of the order of $10^7$-$10^8$ A/cm$^2$, which is the highest value reported for FBS [10]. These properties suggest that this family can be a strong contender for the magnetic application, however, which demands a robust synthesis method to synthesize high-quality and large amounts of bulk samples. This family has a very narrow growth temperature range (920-980°C) as reported for single crystal [8] and polycrystal [9] CaKFe$_4$As$_4$. During the growth process, CaFe$_2$As$_2$ and KFe$_2$As$_2$ generally appear as impurity phases that are more stable than the 1144 phase. However, recent study depict the low temperature synthesis (~700°C) by suggesting the effect of kinetic factors rather than to thermodynamic stability [11]. It is well known that the phase purity and, well-connected grain boundaries are both necessary to observe high superconducting properties, as also reported for 1144 [8-9], which is highly challenging to achieve by CSP [9]. The synthesis of CaKFe$_4$As$_4$ has been reported using a spark plasma sintering (SPS) technique [12] which achieved an onset transition temperature ($T_c^{onset}$) of 35 K with $\Delta T$~2 K and a calculated $J_c$ of 8.1 × 10$^4$ A/cm$^2$ at 5 K and 0 T [12]. This $J_c$ value is slightly higher than that (2.5 × 10$^4$ A/cm$^2$) of 1144 prepared by CSP [9], but both methods have nearly the same sample density (~96-98%). Very few studies have been reported based on the high-pressure growth of FBS, which shows the improved superconducting properties and crystal sizes [13-16]. However, we need more devoted work in this direction. Until now, there has been no report based on the high-pressure synthesis of 1144 polycrystalline and single crystals. This is our main motivation behind this study.

Generally, two types of pressure techniques are used to grow FBS [16-17]: 1) Solid-medium pressure technique: It has a limited sample space of ~0.5 cm$^3$, and there are high chances of the samples being contaminated due to the pressure medium. For example: hot-pressing, hot-isostatic pressing (HIP) [18], and diamond anvil cell (DAC) [13-15]. 2) Gas pressure technique: It has several cm$^3$ of sample space [16], and the pressure is generated by the gas. So, there is no possibility of impurities being introduced into the sample due to the pressure medium. On this basis, we have decided to use HP-HTS for the growth of 1144. Our HP-HTS system is capable of generating gas pressures reaching approximately 1.8 GPa within a cylindrical chamber equipped with a single or multi-zone furnace capable of achieving temperatures of up to 1700°C [16]. In this paper, we have prepared polycrystalline CaKFe$_4$As$_4$ by the HP-HTS method. These samples were well characterized by XRD, microstructure, transport, and magnetic measurements. The superconducting properties are enhanced by the high-pressure techniques compared to the parent compound prepared by CSP. Our study confirms that this pressure synthesis can play an important role in obtaining high-quality 1144 with the improved superconducting properties.

**II. EXPERIMENTAL**

Polycrystalline CaKFe$_4$As$_4$ (parent) was prepared by solid-state reaction method using the high-quality precursors: Ca (purity: 99.5%), K (purity: 99.95%), Fe (purity: 99.99%), and As (purity: 99.999%). First, KAs, Fe$_2$As, and CaAs were prepared by using these precursors. In the next step, all these precursors were mixed according to the stoichiometry of CaKFe$_4$As$_4$ and placed into a Ta-tube, which was sealed into an argon atmosphere using an ARC melter. A two-step process was applied by CSP, as mentioned in our previous study [9].

In our recent study, we have optimized the high gas pressure growth of Fe(Se,Te) (11) bulks by preparing many samples under various growth pressures (0-1 GPa), which suggests that 0.5 GPa for 1 hour can be suitable growth conditions for the 11 family [17] to improve the superconducting properties as well as the sample quality. Hence, these optimized growth conditions (0.5 GPa, 1 hour) were applied for the preparation of 1144 bulks. The prepared 1144 bulks by CSP were used for HP-HTS [17] as an *ex-situ* process, where the samples were heated at 500°C and 0.5 GPa for one hour in an open or closed Ta-tube. Here, two samples were prepared by HP-HTS: First sample, *i.e.*, HIP_1: the parent 1144 bulks was sealed in a Ta-tube by using an ARC melter under an inert atmosphere and heated at these optimized conditions. Second sample, *i.e.*, HIP_2: the parent 1144 was placed into the pressure chamber in an open Ta-tube and heated at 500°C for 1 hour at 0.5 GPa. In the next step, the same sample was sealed in a Ta-tube using an ARC melter and heated at the same pressure, temperature, and time.

Structural characterization has been done by the Rigaku smartLab 3 kW diffractometer with Cu-K$\alpha$ radiation through XRD measurements. ICDD PDF4+2021 standard diffraction patterns database and Rigaku's PDXL software were applied to perform the profile analysis and the calculation of lattice parameters. Microstructural analysis and elemental mapping were performed by a Zeiss Ultra Plus field-emission scanning electron microscope equipped with energy-dispersive X-ray analysis (EDAX) by Bruker mod. Quantax 400 with an ultra-fast detector. Transport properties were characterized by a four-probe method attached to a closed-cycle refrigerator (CCR), through which we can go up to 7 K in zero magnetic field. Magnetic measurements were performed by vibrating-sample magnetometry (VSM) attached to the Physical Property Measurement System (PPMS) in the temperature range of 5-40 K and the magnetic field up to 9 T.

## III. RESULTS AND DISCUSSION

To confirm the phase purity of these prepared 1144 bulks, we have performed XRD, microstructural, and elemental mapping on these samples. Fig. 1 shows the measured XRD pattern for these three samples: Parent, HIP_1, and HIP_2. XRD analysis confirms the tetragonal phase with space group *I4/mmm* (ThCr$_2$Si$_2$-type structure) similar to previous reports [7-10]. Parent and HIP_1 have a clean 1144 phase and no impurity phase was observed, however, a small amount of CaFe$_2$As$_2$ phase is detected for HIP_2. The lattice parameters are $a$ = 3.872(3) Å, $c$ = 12.851(2) Å for the parent compound, $a$ = 3.863(5) Å, $c$ = 12.852(3) Å for HIP_1, and $a$ = 3.852(1) Å, $c$ = 12.863(2) Å for HIP_2, which are almost the same as that reported previously [7][9]

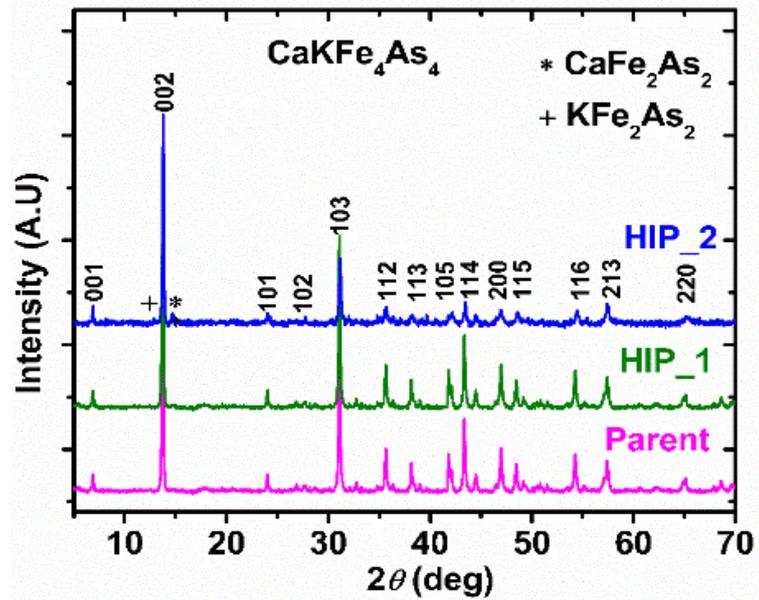

**Fig. 1**. Powder X-ray diffraction (XRD) pattern of CaKFe$_4$As$_4$ synthesis by CSP and HP-HTS method.

In order to understand the microstructural analysis, The backscattered image (BSE) images of these samples are shown in Fig. 2, whereas the elemental mapping is added as Fig. S1 in the supplementary file. Light gray and black contrasts are observed in Fig. 2, corresponding to the 1144 phase and pores, respectively. The parent and HIP_1 have almost homogeneous microstructures, as also supported by their elemental mapping (Fig. S1). HIP_1 has many well-connected grain boundaries (GBs) and seems more compact compared to the parent compound. However, many pores do exist in the parent compared to HIP_1. In the case of HIP_2, it appears that grain connections are reduced rapidly and pore sizes are increased, as shown in Figs. 2(c) and 2(f). The observed white contrast on the sample surface (Fig. 2(c)) could be due to the oxidation of the observed $CaFe_2As_2$ phase, as confirmed by the XRD, which suggests that this sample is not very stable in the air. A recent study [19] has also reported the oxidation of the 1144 phase. The inhomogeneity of the HIP_2 sample is further supported by its elemental mapping (Fig. S1). By considering the theoretical density of 5.22 g/cm$^3$ [9], we have calculated the sample density for the parent compound and HIP_1, which is around 66% and 77%, respectively. This improved density agrees well with the observed BSE images of these samples. However, we were unable to calculate the precise density of HIP_2 because of its small size and irregular shape.

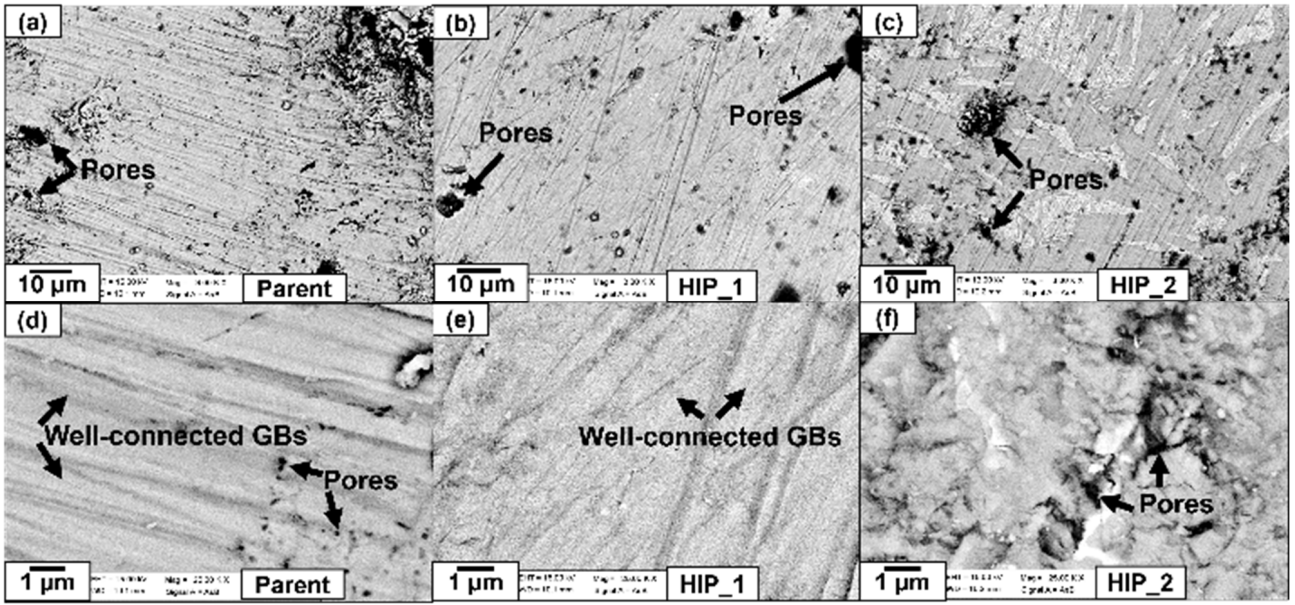

**Fig. 2.** Back-scattered electron (BSE) images for **(a),(d)** the parent; **(b),(e)** HIP_1 and **(c), (f)** HIP_2.

The resistivity behavior of these samples is depicted in Fig. 3(a). All samples showed a linear decrease in the resistivity, and the superconductivity was observed below 40 K. In one batch of the samples, some samples depict a kink in the normal state resistivity around 170 K, as also reported for 1144 single crystals [8,20], which could be due to the magnetic-structural transition of $CaFe_2As_2$. Supplementary Fig. S2 depicts the resistivity behaviors of another sample piece from the same batch, where almost no kink was observed around that temperature. Interestingly, the room-temperature resistivity is around 0.5 mΩ-cm for the parent, which is the same as the previous report [9]. The normal state resistivity is reduced for HIP_1, which might be due to the increased sample density and the observed well-connected grain boundaries, as clear from the microstructural analysis. The presence of numerous pores and the impurity phase in the HIP_2 sample may be the cause of its higher resistivity when compared to HIP_1 and the parent.

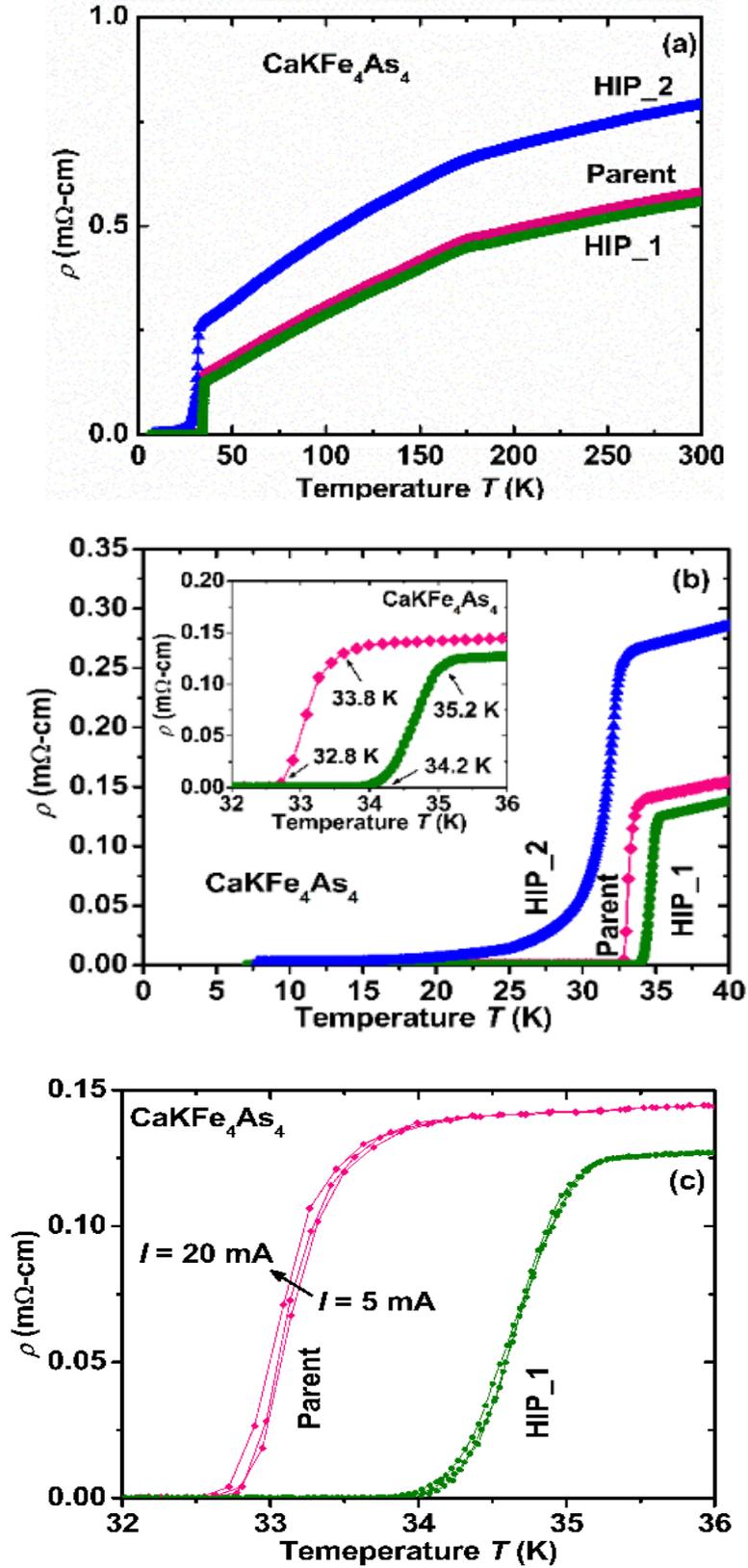

**Fig. 3. (a)** The variation of resistivity ($\rho$) with temperature in zero magnetic fields **(b)** Low-temperature variation of $\rho$ with temperature **(c)** The temperature dependence of the $\rho$ measured under different currents ($I$ = 5, 10 and 20 mA).

To confirm the superconducting transition temperature, the low-temperature resistivity behaviors are depicted in Fig. 3(b). The parent compound has a superconducting transition of ($T_c^{onset}$) 33.8 K with $\Delta T$~1 K, whereas the same superconducting transition width is observed with the enhanced $T_c^{onset}$ value of 35.2 K for HIP_1, which is comparable to that of 1144 single crystal [8],[10]. The sharper transition suggests the high quality of these bulks. The onset and offset transitions by considering 90% and 10% of the normal state resistivity are shown in the inset of Fig. 3(b). The sample HIP_2 depicts almost the same onset transition temperature as the parent compound but has a very broad transition, i.e., the transition width $\Delta T$~12 K, which could be due to weak grain connections as well as the existence of many pores and the impurity phase [2],[21] as discussed above with the microstructural analysis.

To understand the effect of intergrain and intragrain behavior, the temperature dependence of the resistivity for the parent compound and HIP_1 bulks is shown in Fig. 3(c) for various currents ($I$ = 5, 10, and 20 mA). The onset transition is generally related to intragrain behaviors, whereas the offset transition represents the intergrain behaviors [21-22]. The broadening of the onset and offset transition can be seen easily with the effect of various currents for the parent, but HIP_1 sample has almost no broadening. It suggests that HIP_1 sample has better grain connectivity than the parent compound and supports the microstructural analysis.

To confirm the Meissner effect of these samples, we have measured the diamagnetic magnetic moment in the presence of the magnetic field of 20 Oe in zero-field-cooling (ZFC) and field-cooling (FC) mode for the parent and HIP_1 sample. The temperature dependence of the magnetic susceptibility ($\chi$) is shown in Fig. 4(a) for these samples, which depicts an almost full superconducting volume fraction (~96%) and confirms the pure phase formation of 1144. The parent sample illustrates the superconducting transition at 33.3 K as similar to the resistivity measurements. The sample prepared by high-pressure synthesis, i.e., HIP_1, has an onset transition of 34.5 K, which is higher and sharper than the parent sample, suggesting better grain connections [2]. A two-step-like transition is observed for these samples, Fig. 4(a), suggesting a weak-link behavior as well known for FBS [22] and could be due to their lower sample density compared to the previous reports [9][12].

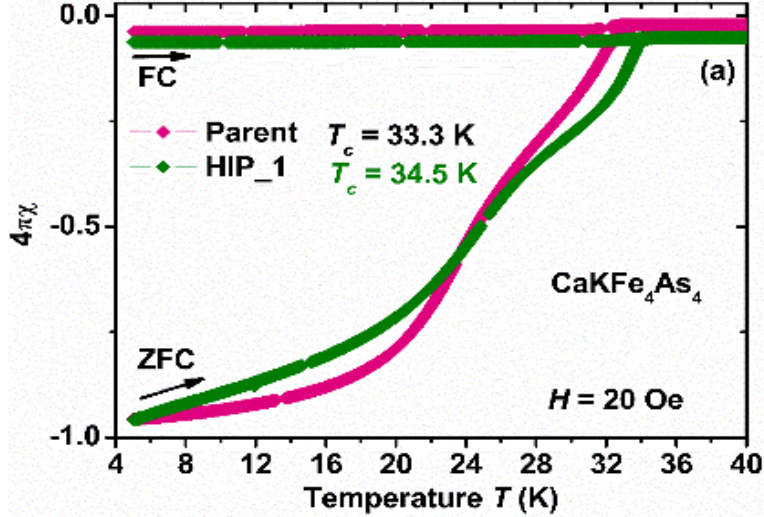

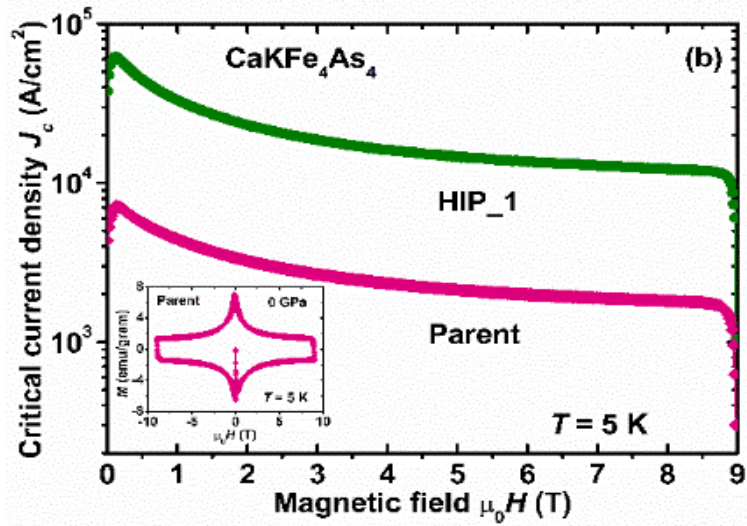

**Fig. 4.** **(a)** The variation of magnetic susceptibility ($\chi$) with temperature in 20 Oe magnetic field **(b)** The magnetic field dependence of $J_c$ for the parent and HIP_1 sample. The inset figure shows the magnetic hysteresis loop (*M-H*) at 5 K for the parent sample.

The critical current density ($J_c$) is an important parameter for practical applications [1]. The magnetic hysteresis loop at 5 K was measured for the parent and HIP_1 sample up to the magnetic field of 9 T. The inset of Fig. 4(b) indicates the hysteresis loop for the parent, which is similar to the previous reports [8-10]. By using the magnetic hysteresis loop, we have calculated the $J_c$ by applying the Bean model $J_c = 20\Delta m / Va(1-a/3b)$ [23], where $\Delta m$ is the difference of magnetization when sweeping the magnetic field up and down, *a* and *b* are the short and long edges of the sample ($a < b$), and *V* is the volume of the sample. The rectangular-shaped samples were used for the hysteresis loop measurements. Using the sample dimensions, the $J_c$ is calculated for the parent sample and HIP_1, as depicted in Fig. 4(b). HIP_2 sample was very small and had an improper shape, so we were unable to calculate the $J_c$, which can be expected to be lower than other samples due to the presence of many pores and impurity phases. The parent sample exhibits a similar magnetic dependence behavior of $J_c$, as reported for 1144, albeit its $J_c$ value ($7.5 \times 10^3$ A/cm$^2$, 0 T) is slightly lower than that of the previous studies [9][12]. It suggests the reduced pinning centers in this parent sample, which could be possible due to its low sample density compared to the reported 1144 bulks [9]. By the high-pressure synthesis method, the $J_c$ value is enhanced by one order of magnitude compared to the parent sample; however, its behavior is the same as that of the parent sample. This $J_c$ enhancement could be due to the improvement of material density, grain connections, and the appropriate pinning centers as reported for $MgB_2$ [24] and FBS [25-26], suggesting that high-pressure synthesis works well to improve the intergrain connections and the pinning properties of 1144.

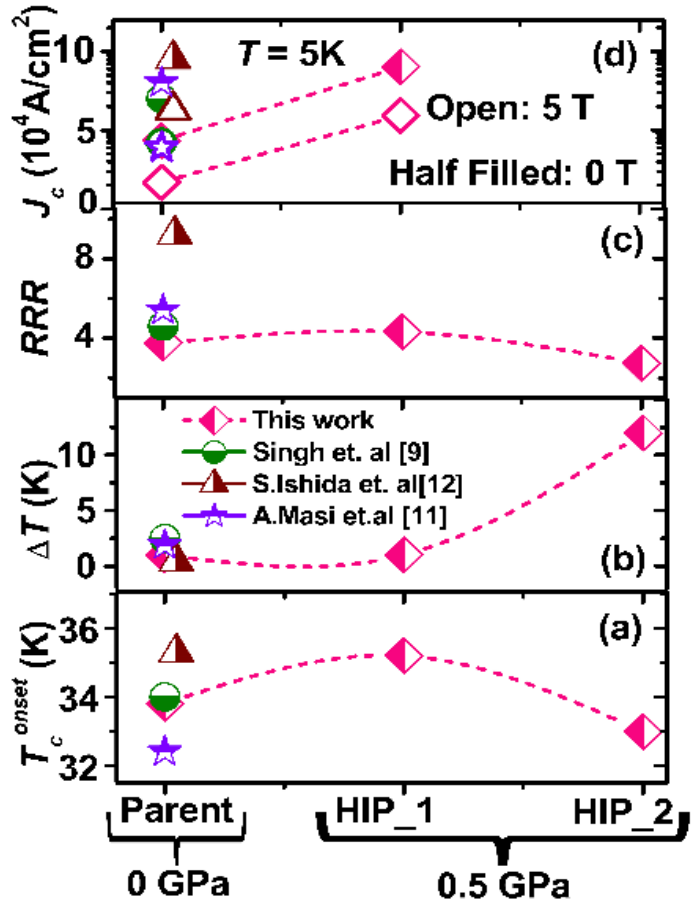

**Fig. 5.** The variation of (a) the onset transition ($T_c^{onset}$) (b) transition width ($\Delta T$) (c) the residual resistivity ratio ($RRR = \rho_{300\,K}/\rho_{40\,K}$) (d) the critical current density ($J_c$) at 0 T and 5 T at 5 K for the samples prepared at different pressure for CaKFe$_4$As$_4$ bulks. The data are also included from the previous reports for 1144 bulks.

The main findings of this paper are depicted in Fig. 5(a)-(d) with respect to the previous reports at ambient pressure [9,11] and by SPS [12]. The $T_c^{onset}$ value reached a maximum for HIP_1 as reported by SPS [12] and its transition width ($\Delta T = 1$ K) is the same as that of our parent compound and other reports [9,12]. HIP_1 has a slightly improved *RRR* value (4.3) than that (3.7) of the parent sample, suggesting a homogeneous and good-quality of HIP_1 sample with respect to our other samples. This value lies in the same range as the reported *RRR* value for 1144 [9,11] prepared by CSP, however, it is lower than that of 1144 prepared by SPS [12] (Fig. 5(c)). The variation of $J_c$ is depicted in Fig. 5(d) with respect to the previous reports [9,11-12]. Interestingly, HIP_1 has enhanced the Jc value in the whole magnetic field up to 9 T compared to the parent samples, and this value is almost the same as a value reported for 1144 prepared by CSP [9] and SPS [12]. All these results suggest that $CaKFe_4As_4$ prepared at 0.5 GPa by HP-HTS into an open Ta-tube exhibits high superconducting properties with the improved sample quality, but still has a low sample density compared to the previous reports [9,12]. So, it appears that high-pressure synthesis has been successful for the 1144 family; nevertheless, additional studies in this area are required to further enhance its superconducting characteristics.

## IV. CONCLUSIONS

A high gas pressure growth of $CaKFe_4As_4$ is employed to study the effect on its superconducting properties. The samples prepared through HP-HTS in an open Ta-tube have improved grain connectivity, sample density, and homogeneous distribution of the constituent elements, whereas the second sample (HIP_2) prepared through HP-HTS in an open Ta-tube and then sealed in a Ta-tube i.e. long heating time under high pressure, has reduced sample quality and superconducting properties. The best 1144 bulks by high-pressure growth have shown a high $T_c$ value of 35.2 K and also a high $J_c$ value of the order of $6 \times 10^4$ A/cm$^2$, which could be due to the improved sample density, good grain connectivity, and effective pinning centers. Our study confirms that high gas pressure synthesis works well to improve the superconducting properties and sample quality of 1144. Also, it demands more research work to further improve the physical properties of 1144 materials by considering various growth parameters such as pressure, temperature, and time.

## V. ACKNOWLEDGMENT

This research was funded by National Science Centre (NCN), Poland, grant number "2021/42/E/ST5/00262" (SONATA-BIS 11). S.J.S. acknowledges financial support from National Science Centre (NCN), Poland through research Project number: 2021/42/E/ST5/00262.

**(i)**

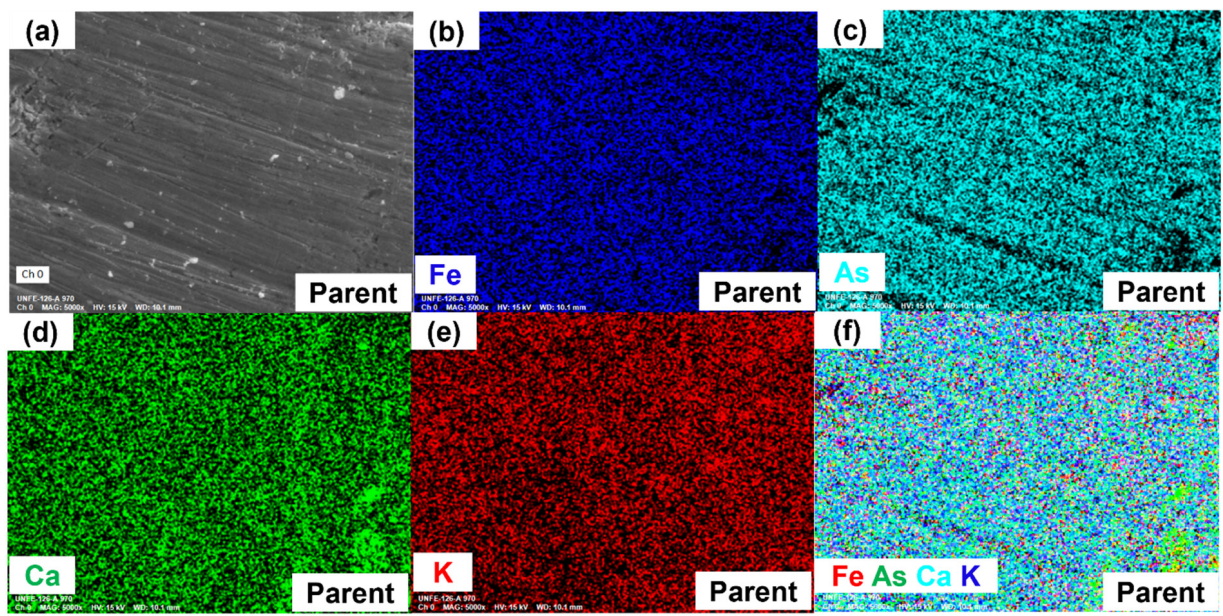

**(ii)**

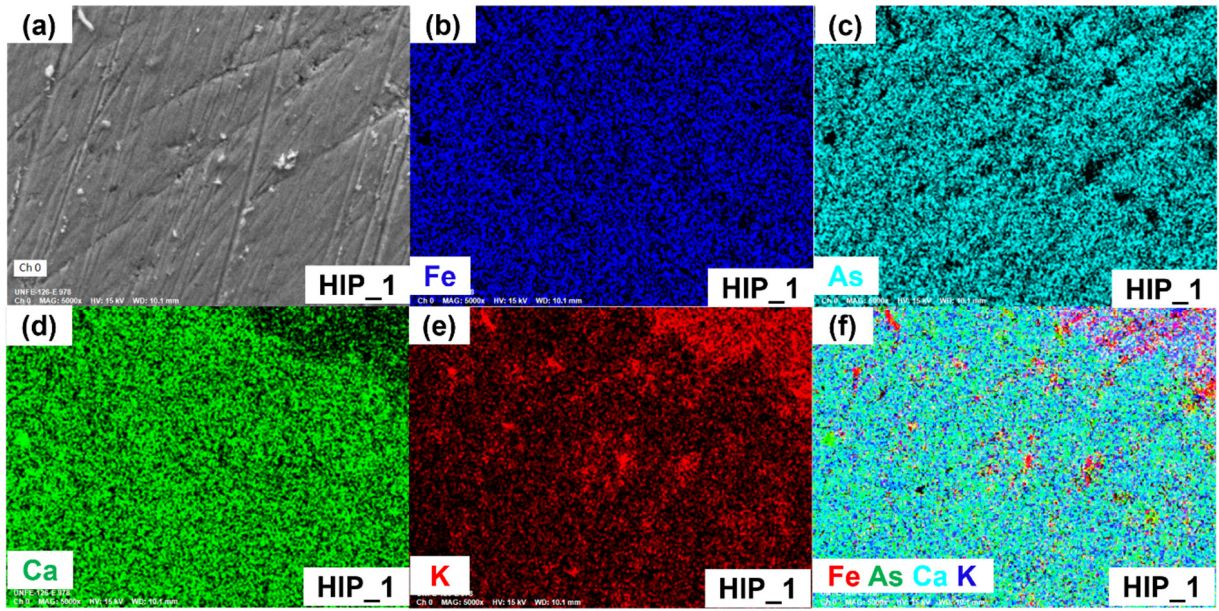

**(iii)**

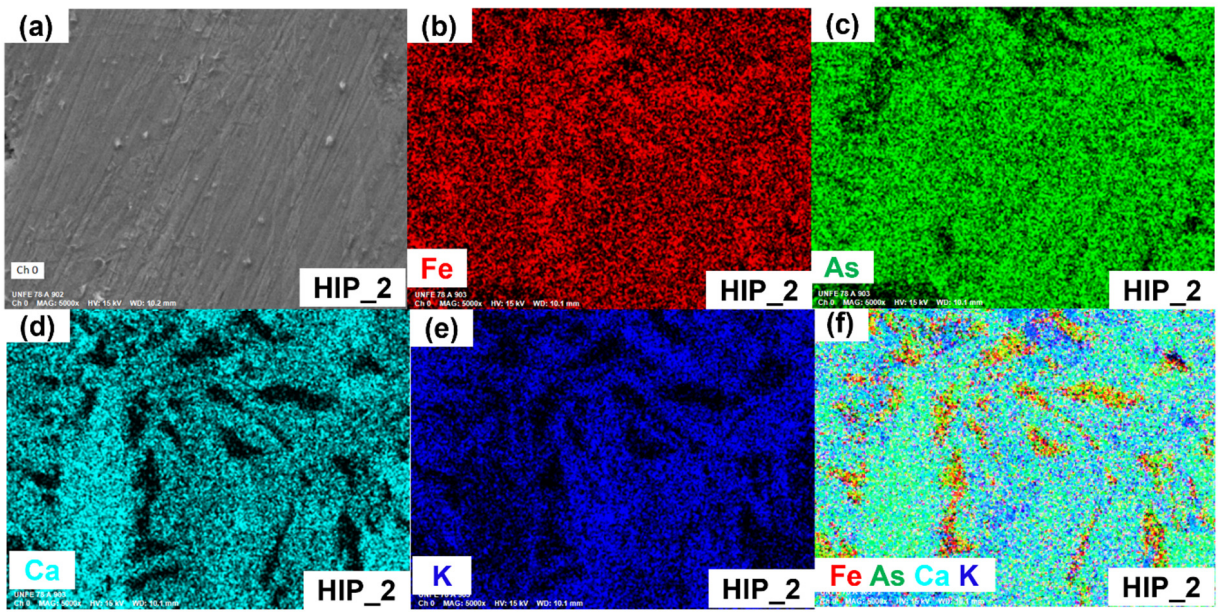

**Fig. S1**: Elemental mapping of all constituent elements for the sample: **(i)** Parent; **(ii)** HIP_1 and **(iii)** HIP_2.

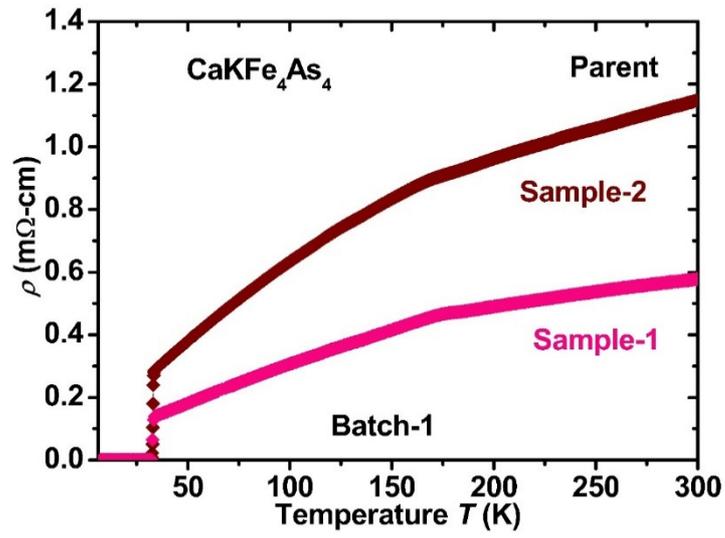

**Fig. S2**: The variation for resistivity ($\rho$) with temperature for two rectangular pieces (Sample-1 and Sample-2) of the parent sample from the same batch (**Batch-1).**